\documentclass[3p,twocolumn,number]{elsarticle}

\biboptions{sort&compress}

\usepackage{latexsym}
\usepackage{amsfonts}
\usepackage{amsmath}
\usepackage{amssymb}
\usepackage{booktabs}
\usepackage{mathrsfs}
\usepackage{caption}
\usepackage{color}
\usepackage{graphicx}
\usepackage{mathptmx}      
\usepackage{bm}
\usepackage{enumerate}
\usepackage{subfigure}
\usepackage{listings}
\usepackage{makecell}
\usepackage{grffile}
\usepackage{multirow}
\usepackage{quantikz}
\usepackage{siunitx}
\usepackage{mathtools}
\usepackage[normalem]{ulem}
\usepackage{pdfpages}
\usepackage{float}
\usepackage{lineno}
\usepackage{xcolor}
\usepackage{xurl}
\usepackage{hyperref}
\hypersetup{
  colorlinks=true,
}
\usepackage{arydshln}

\definecolor{codegreen}{rgb}{0,0.6,0}
\definecolor{codegray}{rgb}{0.5,0.5,0.5}
\definecolor{codepurple}{rgb}{0.58,0,0.82}
\definecolor{backColour}{rgb}{0.95,0.95,0.92}
\definecolor{mGreen}{rgb}{0,0.6,0}
\definecolor{mGray}{rgb}{0.5,0.5,0.5}
\definecolor{mPurple}{rgb}{0.58,0,0.82}
\definecolor{backgroundColour}{rgb}{0.97,0.97,0.97}
\definecolor{mBlue}{rgb}{0.1,0.1,0.8}

\lstdefinestyle{CStyle}{
    backgroundcolor=\color{backColour},
    commentstyle=\color{mGreen},
    keywordstyle=\color{mBlue}\bfseries,
    numberstyle=\tiny\color{mGray},
    stringstyle=\color{mPurple},
    basicstyle=\ttfamily\footnotesize,
    breakatwhitespace=false,
    breaklines=true,
    captionpos=b,
    keepspaces=true,
    numbers=none,
    numbersep=5pt,
    showspaces=false,
    showstringspaces=false,
    showtabs=false,
    tabsize=4,
    language=C++,
    morecomment=[l][\color{black}]{\#}
}

\DeclareSIUnit\flop{FLOP}
\DeclareSIUnit\cycle{\textsc{Cycle}}
\DeclareSIUnit[per-mode=symbol]\floppersec{\flop\per\second}
\DeclareSIUnit[per-mode=symbol]\flopperjoule{\flop\per\joule}
\DeclareSIQualifier{\doubleprecision}{FP64}
\DeclareSIQualifier{\singleprecision}{FP32}
\DeclareSIQualifier{\halfprecision}{FP16}
\DeclareSIQualifier{\theoretical}{th}
\DeclareSIUnit\pixel{px}

\definecolor{fzjblue}{RGB}{2,61,107}
\definecolor{fzjlightblue}{RGB}{173,189,227}
\definecolor{fzjgray}{RGB}{235,235,235}
\colorlet{fzjgrey}{fzjgray}
\definecolor{fzjred}{RGB}{235, 95, 115}
\definecolor{fzjgreen}{RGB}{185, 210, 95}
\definecolor{fzjyellow}{RGB}{250, 235, 90}
\definecolor{fzjviolet}{RGB}{175, 130, 185}
\definecolor{fzjorange}{RGB}{250, 180, 90}
\definecolor{fzjblack}{RGB}{0,0,0}
\definecolor{fzjwhite}{RGB}{255,255,255}

\newcommand{\nvidia}{NVIDIA}

\usepackage{tikz}
\newcommand*\circled[2][2pt]{\tikz[baseline=(char.base)]{
            \node[shape=circle,inner sep=#1,fill=gray, text=white] (char) {#2};}}

\begin{document}
\begin{frontmatter}

\title{%
Universal Quantum Computer Simulation of 50 Qubits on Europe’s First Exascale Supercomputer Harnessing Its Heterogeneous CPU-GPU Architecture%
}
\cortext[cor1]{Corresponding author: v.mehta@fz-juelich.de
}
\address[FZJ]{J\"ulich Supercomputing Centre, Forschungzentrum J\"ulich, D-52425 J\"ulich, Germany}
\address[NVIDIA]{NVIDIA, W\"urselen, Germany}

\author[FZJ]{Hans De Raedt}
\author[NVIDIA]{Jiri Kraus}
\author[FZJ]{Andreas Herten}
\author[FZJ]{Vrinda Mehta\corref{cor1}}
\author[FZJ]{Mathis Bode}
\author[NVIDIA]{Markus Hrywniak}
\author[FZJ]{Kristel Michielsen}
\author[FZJ]{Thomas Lippert}



\begin{abstract}
    We have developed a new version of the high-performance Jülich universal quantum computer simulator (JUQCS-50) that leverages key features of the GH200 superchips as used in the JUPITER supercomputer, enabling  simulations of a 50-qubit universal quantum computer for the first time.
    JUQCS-50 achieves this through three key innovations: (1) extending usable memory beyond GPU limits via high-bandwidth CPU-GPU interconnects and LPDDR5 memory; (2) adaptive data encoding to reduce memory footprint with acceptable trade-offs in precision and compute effort; and (3) an on-the-fly network traffic optimizer. These advances result in an {\color{black}16.6}-fold speedup over the previous 48-qubit record on the K computer.
\end{abstract}


\end{frontmatter}
{\color{black}
\section{Introduction}
Quantum computing promises transformative advances in cryptography, materials science, quantum chemistry, and artificial intelligence~\cite{gill2020quantum}. However, current quantum hardware remains limited by noise, gate infidelity, and restricted circuit depth, making large-scale, reliable execution infeasible. As a result, simulation of quantum circuits on digital computers remains essential for algorithm development and benchmarking.

Simulation of quantum circuits on digital computers is computationally demanding, as both memory usage and runtime scale exponentially with the number of qubits~\cite{Nielsen2010}. Recent progress in high-performance quantum simulators has begun to mitigate these challenges by exploiting massively parallel architectures. JUQCS-50 is a high-fidelity universal quantum computer simulator that demonstrates near-linear scalability in elapsed time with increasing qubit count, enabling simulations at unprecedented scale.

Leveraging the JUNIQ (Jülich Unified Infrastructure for Quantum Computing) platform and the JUPITER supercomputer, JUQCS-50 supports the simulation of quantum circuits with up to 50 qubits for the first time. This capability enables accurate benchmarking of algorithms such as the Variational Quantum Eigensolver (VQE) and the Quantum Approximate Optimization Algorithm (QAOA), well beyond the reliable operating regime of current quantum processors.

{\color{black}
Although some quantum processors now exceed 50 qubits, current devices still suffer from substantial noise and limited gate fidelities, which severely restrict the reliable circuit size and depth. For example, a simple 4‑bit adder (depth 10) on an IQM chip achieves only a \qtyrange{70}{80}{\percent} success rate, and performance degrades rapidly for larger circuits (a 50‑qubit adder requires 1001 gates and depth 101). Even recent breakthroughs, such as the demonstrations of surface‑code quantum error correction below threshold~\cite{Acharya2024}, where increasing the code distance yields exponentially improved logical lifetimes, highlight that today’s hardware is only beginning to enter the regime where scalable fault‑tolerant computation becomes feasible. In contrast, classical simulations of the same circuits produce numerically exact results, unaffected by hardware noise or decoherence.
}

This work makes the following contributions:
\begin{itemize}
\item Demonstration of the first universal 50-qubit quantum computer simulation, using exascale capabilities.
\item High-performance system design achieving efficient execution on \num{16384} GH200 superchips using hybrid memory and optimized communication.
\item Adaptive byte encoding combining low-precision arithmetic with FP64 to reduce memory footprint and improve performance.
\item Comprehensive performance evaluation, including weak and strong scaling studies using Hadamard circuits. 
\end{itemize}}
The remainder of this paper is organized as follows.
The \hyperref[sec:problem]{next Section} briefly discusses the computational problem. \hyperref[STATE]{Section~\ref*{STATE}} provides an overview of the current state of the art in quantum computer simulation techniques. \hyperref[sec:innovations]{Section~\ref*{sec:innovations}} describes the methods and optimizations implemented in JUQCS-50 that enable simulations at the 50-qubit scale. \hyperref[sec:perfmeas]{Section~\ref*{sec:perfmeas}} introduces the performance metrics used to evaluate JUQCS-50, and \hyperref[sec:PerfResults]{Section~\ref*{sec:PerfResults}} presents the corresponding performance results. In \hyperref[sec:adder]{Section~\ref*{sec:adder}}, we discuss an application of JUQCS-50 to adder circuits, including both the approach and results. Finally, \hyperref[sec:Summary]{Section~\ref*{sec:Summary}} concludes the paper with a summary of the key findings and perspectives for future work.

\section{Problem statement}\label{sec:problem}
A universal quantum computer~\cite{Nielsen2010} simulator requires storage for
the full state vector (wave function) describing the state of the quantum computer~\cite{Nielsen2010}.
Using the FP64 representation, storing the (complex-valued) state vector for an $N$-qubit quantum computer
requires $2^{N+4}$ bytes of memory, $2^{4}$ accounting for the number of bytes required to store an element of the state vector.
For example, representing the state of a $N=32$ qubit quantum computer requires \qty{64}{\gibi\byte} of memory in FP64 precision.

A quantum program is a sequence of quantum gates~\cite{Nielsen2010}, each of which modifies the state vector. In general, each gate affects all the elements of the state vector. Simulating a single-qubit quantum gate amounts to multiplying all $2^{N/2}$ disjoint pairs of state vector elements by a $2\times2$ matrix (corresponding to the quantum gate), the choice of pairs being determined by the particular qubit on which the quantum gate is applied. Similarly, simulating a two-qubit quantum gate amounts to multiplying all $2^{N/4}$ disjoint quadruples of state vector elements by a $4\times4$ matrix (corresponding to the quantum gate), the choice of quadruples being determined by the particular qubits on which the gate is applied. As all these matrix-vector multiplications involve disjoint elements only, these operations can be carried out in parallel. Depending on the qubit(s) addressed, the pairs (quadruples) of vector elements may be located very far apart in memory. For instance, for a single-qubit gate acting on one of the qubits $k=0,\ldots,N-1$, the two indices of a pair differ by $2^{k}$. This is a characteristic feature of a gate-based quantum computer simulator and is of prime importance when the state vector becomes so large that the memory storing all its elements has to be distributed over several processing entities--NVIDIA Grace Hopper GH200 \emph{superchips}, in our case (see \autoref{sec:innovations}).

In the distributed-memory setting, it is convenient to use as the superchip index, the integer representation of as many of the high-order bits of the element address as needed~\cite{RAED07x}. Let $N'$ denote the number of qubits for which the state vector fits into the memory of one GH200 superchip.
Then the number of superchips $n$ required to simulate an $N$ qubit quantum computer and the number of state vector elements $L$ per superchip 
are given by $n=2^{N-N'}$ and $L=2^{N'}$, respectively. Performing a single-qubit gate on qubit with index $0\le j< N'$, requires no exchange of data between superchips because each pair of amplitudes that need to be updated resides in the memory of the same superchip, and all pairs can be updated in parallel.The same holds for two-qubit gates involving qubits $0\le j,k< N'$.
These are the embarrassingly parallel cases: all superchips can perform the necessary calculations independently without any communication between the superchips. In contrast, for a single-qubit gate on qubit $N'\le j< N$ (or a two-qubit gate
with its target qubit~\cite{Nielsen2010} index exceeding $N'-1$), superchips have to exchange data.

Focusing on the single-qubit case for simplicity, it follows that
half of the state vector elements stored in each superchip need to be exchanged~\cite{RAED07x}.
This redistribution of elements always involves pairs of superchips, the indices of these superchips
depending on the qubit that is being considered.
For two-qubit gates, three-quarters of the number of elements need to be transferred between pairs of superchips.
\begin{figure*}
    \centering%
    \includegraphics[width=0.95\linewidth]{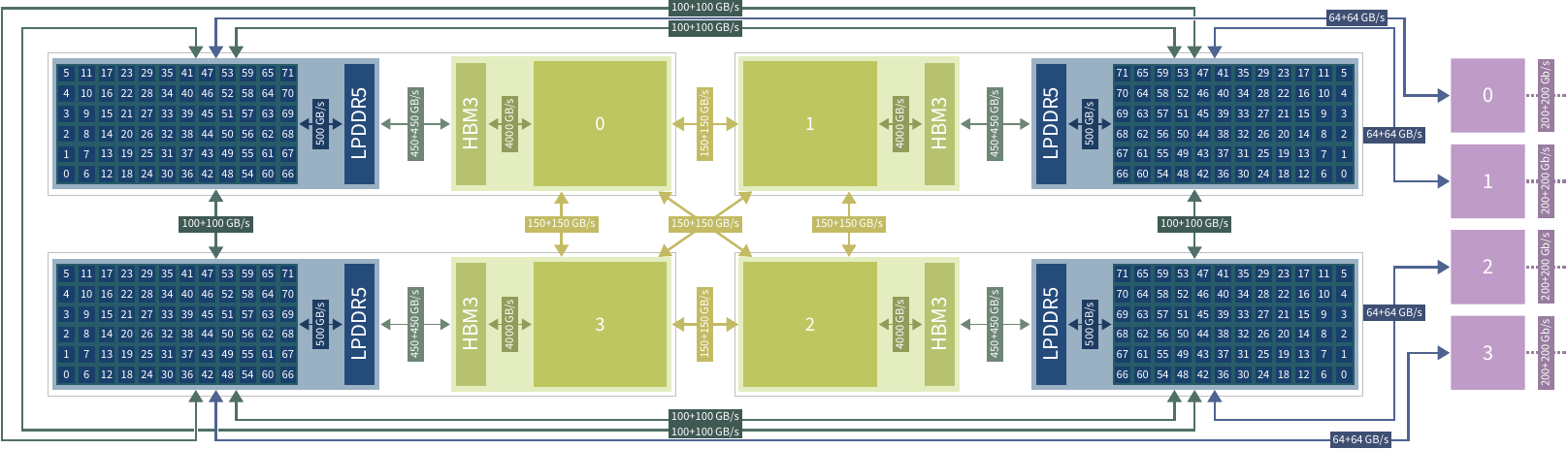}
    \caption{Overview of JUPITER's quad GH200 node design with included technology and bandwidths. Each node contains four GH200 superchips, each comprising a tightly integrated pair {\color{black}of CPU (blue) and GPU (green)}; see \autoref{sec:innovations}. {\color{black}Each processor includes its main memory (LPDDR5 and HBM3, respectively); CPUs have individual cores shown. Purple blocks indicate network interfaces. Lines between node blocks indicate  available connections, with respective bandwidths superimposed.}}
    \label{fig:jupiter-node-diagram}
\end{figure*}

As an example, consider simulating a $N=40$ qubit quantum computer on the  JUPITER supercomputer, requiring a total
of $2\times8\times2^{40}/2^{30}=\qty{16384}{\gibi\byte}$ to store all state vector elements in FP64 precision.
Because of the inherent restriction to powers of two, characteristic of qubit systems, each GH200 GPU with \qty{96}{\gibi\byte} memory (see \autoref{fig:jupiter-node-diagram} for a diagram of a node) can hold the equivalent of $2^{32}$ of these state vector elements (\qty{64}{\gibi\byte}).
This implies that with each single-qubit gate applied to a qubit with index $j>31$, \qty{16384}{\gibi\byte} of data traverses through the network fabric
before the gate operation can actually be carried out.
Even for the fastest communication networks that are available today, it is clear that as the number of qubits $N$ increases, the data exchange between superchips, when not performed efficiently, may become a very time-consuming and limiting part of the simulation. In summary,
\begin{enumerate}
\item
Required memory and compute time increase exponentially with the number of qubits $N$.
\item
The number of floating point operations per gate is given by $a 2^N$, where $a$ is a factor
that depends weakly on the kind of gate.
\item
For all gates involving qubits with indices smaller than $N'$, there is no communication between superchips.
\item
For a single-qubit gate on the qubit with index $N'\le j< N$, all superchips
have to perform a send-receive operation, exchanging $2^{N'}/2$ elements of the state vector.
\item
For a two-qubit gate involving a target qubit with index in the interval {\color{black}$[N',N)$}, all pairs of superchips
have to perform a send-receive operation, exchanging $3\times2^{N'}/4$ elements of the state vector
\end{enumerate}

\section{Current State of the Art}\label{STATE}

As explained above, dealing with the exponential growth of computational resources as the number of qubits increases is the main challenge of simulating quantum systems.
Various techniques have been developed to address this challenge,
each with its own strengths and weaknesses.

\subsection{State Vector Simulation}
Conceptually, state vector simulation is the most straightforward method,
representing the full quantum state as a complex-valued vector of dimension
\(2^N\), where \(N\) is the number of qubits.

This method allows for exact simulations but is resource-intensive.
Examples are IBM Qiskit's AerSimulator~\cite{Qiskit}, Google Cirq~\cite{Cirq}, Eviden's Qaptiva~\cite{Qaptiva}, and JUQCS~\cite{RAED19a,Willsch2020}.
Memory and processing constraints currently limit the applicability
of the three former examples to problems of 30--41 qubits.
JUQCS goes significantly beyond this: the largest universal quantum computer JUQCS has simulated thus far contained 48 qubits~\cite{RAED19a}.

Just like its predecessors~\cite{RAED07x,RAED19a,Willsch2022}, which have set multiple world records for simulating the largest universal quantum computers~\cite{WEB1,WEB2}, JUQCS-50 is designed with a strong emphasis on portability. It runs seamlessly across a wide spectrum of hardware platforms--from desktop PCs to high-end supercomputers with distributed and shared memory. This flexibility makes JUQCS-50 accessible to a broad user base, enabling both development and large-scale production runs in diverse computing environments. On CPU-based systems, only a Fortran 2003-compatible compiler (e.g., Intel IFX, gfortran, nvfortran) and standard MPI support are required. To utilize \nvidia{} GPUs, JUQCS-50 additionally depends on CUDA-Fortran and CUDA-aware MPI, ensuring efficient execution while maintaining ease of deployment across systems.
Notably, the majority of JUQCS-50 software development was conducted on the JUWELS Booster system~\cite{BOOSTER}, utilizing NVIDIA A100 GPUs and AMD CPUs. This preceded the operational launch of JUPITER and underscores the feasibility of enablement work on previous-generation hardware. 

{\color{black}JUQCS-50 is functionally portable across CPU-only and GPU-accelerated systems (Fortran 2003 + MPI, with optional CUDA-Fortran and CUDA-aware MPI). However, the performance results are hardware-specific, relying on GH200 features such as the high-bandwidth NVLink-C2C CPU–GPU interconnect and unified use of GPU HBM and CPU LPDDR5 memory. The reported performance therefore reflects portability within GH200-like heterogeneous architectures with comparable bandwidth and memory hierarchies.}

In general, to keep the elapsed time of universal quantum computer simulations within reasonable bounds, it is essential to have effective algorithms and hardware to handle the massive amount of data transfers that result from distributing memory over many computational units.

JUQCS-50 addresses all these issues, facilitating the simulation of 50 qubits on the imminent exascale computer JUPITER, deploying GH200 superchips.
JUQCS-50 can exploit new powerful features of the  superchips to enable the simulation of these qubits 
using \num{4096} JUPITER nodes, with a significant reduction of the elapsed time to execute quantum circuits.

\subsection{Tensor Network Simulation}
Tensor networks are mathematical structures that can efficiently represent quantum states if entanglement is limited.
This approach can significantly reduce memory usage for circuits with low-depth or limited connectivity.
Instead of storing the entire state vector,
tensor network methods break down computations into small, manageable parts.
In practice, they are very useful for simulating shallow circuits with limited entanglement and tree-like circuit structures.
One prime, typical example of this approach is the random quantum circuit simulator for the new generation of Sunway Supercomputers~\cite{Liu2021}.
This simulator exploits the particular structure of the random quantum circuits used in Google's quantum supremacy work~\cite{GOOG19} to construct a very efficient scheme to compute
a small fraction of the exponentially large number of amplitudes for systems containing up to 100 qubits~\cite{Liu2021}.
In contrast, JUQCS was used to compare the frequencies of the states produced by the supremacy experiments with the exact simulation of several random quantum circuits employed in the
experiments, up to $N=43$ qubits~\cite{GOOG19}.

{\color{black}
\subsection{Circuit cutting/entanglement-removal approach}
If the number of entangling gates is small, a circuit-cutting strategy based on a Controlled-Z (CZ) 
gate decomposition enables the simulation of circuits with up to 64 qubits using modest computational resources \cite{Chen2018}. 
In this approach, each CZ gate is replaced by a measurement-based decomposition, 
which removes entanglement and splits the original circuit into many independent subcircuits. 
These smaller subcircuits can then be simulated separately, 
and their contributions are recombined to recover the full result. 
This method is most effective for low-depth circuits with sparse CZ gates, 
since the number of subcircuits grows exponentially with the number of decomposed CZ gates. 
In contrast, JUQCS-50 performs a full state-vector simulation and relies on aggressive memory 
and communication optimizations rather than circuit decomposition.
}

\subsection{Dedicated simulation software for Shor's algorithm}
Shor's algorithm for factoring integers is one of the most anticipated applications of quantum computing~\cite{gill2020quantum,Nielsen2010}.
It also provides another instance of dedicated simulation
software executing a particular quantum algorithm for problem sizes that are out of reach for universal quantum computer simulators.
The largest semiprime that has been factored by a universal quantum computer simulator (JUQCS) executing Shor's original algorithm is \num{65531}.
This simulation required simulating a 48-qubit universal quantum computer~\cite{RAED19a}.
In contrast, with a completely different, dedicated implementation of the same algorithm,
employing GPUs and dedicated post-processing of the outcomes of the simulation~\cite{Willsch2023},
the largest semiprime that could be factored
without exploiting prior knowledge of the solution is $\num{549755813701} = \num{712321} \times \num{771781}$~\cite{Willsch2023}.
As in the case of tensor network simulation, this illustrates once more that comparing simulators,
highly tuned to specific problems, with universal quantum computer simulators, is a delicate issue.

\subsection{Summary}
{\color{black}Although specialized simulators can handle larger qubit counts for certain structured circuits, their efficiency depends on exploitable properties such as limited entanglement or algorithm-specific constraints. While alternative methods can outperform JUQCS for particular problems or restricted queries, they do not generalize to unrestricted universal quantum computation. JUQCS-50 instead supports full state vector evolution for arbitrary circuits, enabling general-purpose hardware benchmarking, software verification, and algorithm development without assumptions on circuit structure.}

As a simulator of a universal quantum computer, capable of executing arbitrary quantum circuits composed of universal gates~\cite{Nielsen2010} (and beyond), JUQCS distinguishes itself through unmatched performance and the ability to handle a record number of qubits.

\section{Innovations Realized}\label{sec:innovations}
Consider simulating an $N=50$ qubit quantum computer in FP64 precision on a GH200-based system such as JUPITER. Storing all the elements of the state vector requires $2\times8\times2^{N}/2^{40}=2^{14}=\qty{16384}{\tebi\byte}$ of memory.
In AI terminology,
this quantum computer simulator can be viewed as a $2^{50}-1\approx1024$~trillion (not billion) parameter model
($-1$ because of the normalization of the state vector).

JUPITER\footnote{In this work, \emph{JUPITER} refers to the Booster module of the JUPITER supercomputer. A CPU-centric second module, JUPITER Cluster, will be deployed in the future.} has about 6000 nodes (\num{24000} superchips). Each of the superchips is equipped with \qty{96}{\gibi\byte} of HBM3 (device) and \qty{120}{\gibi\byte} of LPDDR5 (host) memory; \qty{216}{\gibi\byte} in total per GH200 superchip. {\color{black} Because the state vector size must be a power of two, the effective memory available for storing the state vector on a single GH200 superchip is therefore limited to \qty{128}{\gibi\byte}. This memory is equally divided across the GPU and CPU. Memory for buffers (much less than \qty{96}{\gibi\byte}-\qty{64}{\gibi\byte}=\qty{32}{\gibi\byte}) is allocated on the GPU. Again, due to the powers-of-two restriction, the largest number of nodes that can actually be used for JUQCS is 4096 (\num{16384} superchips), yielding in total $\qty{16384}\times\qty{128}{\gibi\byte}=\qty{2048}{\tebi\byte}$,
a factor of eight too small for simulating 50 qubits. }

As previously demonstrated~\cite{RAED19a}, the memory requirement can be reduced by a factor of eight by using a  special form of adaptive byte encoding (see \autoref{sec:byteencoding}),
representing each state vector element by 2 bytes ($2\times\qty{8}{\bit}$) instead of 16 bytes ($2\times\qty{64}{\bit}$, for a complex number in FP64 precision).
This compression reduces precision in general cases (though not in specific ones, such as the gate sequences used for benchmarking in this work)
and increases computation time (see \autoref{sec:byteencoding}).

Employing this technique makes it possible to reach the 50-qubit barrier
with a universal quantum computer simulator running on JUPITER.
A key question, then, is how to let JUQCS-50 compute entirely on the GPU using also the LPDDR5 host memory in its GH200 superchip efficiently within the GPU-based execution with minimal performance impacts. {\color{black} To this end, we adopt a design in which the CPUs act solely as memory hosts, while all computation, including encoding and decoding as well as communication, is performed on the GPUs.}

In conclusion, to push the boundaries of what can be simulated with state-of-the-art hardware such as JUPITER, it is necessary to
\begin{center}
\framebox{
\parbox[ht]{0.9\hsize}{%
\hangafter=1
\begin{enumerate}
\item
maximize the utilization of available memory across GH200 superchips (in powers of two)
\item
minimize the elapsed time spent on communication
\end{enumerate}
}}
\end{center}

In the following, we outline our unique solutions for these requirements. {\color{black}As the software developed is proprietary, the internal algorithmic implementation is not publicly available. The software version, parameters, and workflow applied in this study are documented to ensure reproducibility.}

\subsection{Intra-GH200 communication}

The Grace Hopper superchip is a hardware-coherent GPU-accelerated system where all processors can access all memory with high performance through the \qty{900}{\giga\byte\per\second} NVLink-C2C connection. The most productive approach to use the full memory is through the Unified Memory concept, in which applications interact with one large memory pool. The Operating System (\emph{OS}; Linux) natively supports the GH200 superchip and views both
the CPU and the GPU memory as separate NUMA (\emph{Non-Uniform Memory Access}) nodes. It is thus possible to use native, OS-provided allocators and rely on the OS capabilities to utilize multiple NUMA nodes to create a single, large allocation exceeding the size of the memory of a single NUMA node, usable from both the CPU and GPU parts of GH200. Of course, the CUDA allocation mechanism can also be used. Approximately sorted with increasing programming effort, the options to use GH200's memories are:
\begin{itemize}

    \item[{\footnotesize\circled{1}}] Use \texttt{malloc()}, which does not require any code changes, and rely on OS and Unified Memory Driver heuristics for data placement.
    
    \item[{\footnotesize\circled{2}}] Use \texttt{malloc()}, which does not require any code changes, and guide data placement externally with \texttt{numactl --interleave}    \footnote{Or \texttt{--weighted-interleave} which we could not use due to missing kernel support on the machines.}.
    
    \item[{\footnotesize\circled{3}}] Use \texttt{numa\_alloc\_onnode()} to explicitly place the real and the imaginary parts in different NUMA nodes.

    \item[{\footnotesize\circled{4}}] Use Unified Memory Data Usage Hints of CUDA's Unified Memory API~\cite{CUDADataUsageHintsDoc} to
    explicitly stripe allocations between LPDDR5 and HBM3 memory. {\color{black}\autoref{lst:umdh} shows the code snippet used for this particular implementation.}
    
    \item[{\footnotesize\circled{5}}] Use explicit data movement functions from the CUDA API like \texttt{cudaMemcpyAsync()} to move portions of the data between host and device memories, combined with automatic, on-the-fly optimization of network traffic. {\color{black} The corresponding code is proprietary and is not publically available.}
\end{itemize}
The Unified Memory Data Usage Hints support the Unified Memory Driver with indications of data locality and promise to be a good middle-ground between automated data placement and programming effort. The implementation uses code as outlined in \autoref{lst:umdh}

\begin{lstlisting}[style=CStyle, caption={Core of the Unified Memory Data Usage Hints implementation. {\color{black}
The swap functions update the variables next\_loc and next\_chunksize within the loop}}, label={lst:umdh}]
size_t bytes_remaining = size_in_bytes;
char* psi_r_pos = (char*)psi_r;
char* psi_i_pos = (char*)psi_i;
while (bytes_remaining >= current_chunk_size) {
  cudaMemAdvise_v2(psi_r_pos, current_chunk_size, cudaMemAdviseSetPreferredLocation, current_loc);
  cudaMemPrefetchAsync_v2(psi_r_pos, current_chunk_size, current_loc, 0, 0);
  cudaMemAdvise_v2(psi_i_pos, current_chunk_size, cudaMemAdviseSetPreferredLocation, current_loc);
  cudaMemPrefetchAsync_v2(psi_i_pos, current_chunk_size, current_loc, 0, 0);
  psi_r_pos += current_chunk_size;
  psi_i_pos += current_chunk_size;
  bytes_remaining -=  current_chunk_size;
  std::swap(current_loc,next_loc);
  std::swap(current_chunk_size, next_chunk_size);
}
\end{lstlisting}
to distribute page-size-aligned chunks alternately between
the LPDDR5 and the HBM3 NUMA node iteratively. This mechanism balances concurrent HBM3 and LPDDR5 usage. In the code, \texttt{current\_loc} is initialized with the HBM3 NUMA node of the GPU used by the process, while \texttt{next\_loc}
is initialized with the LPDDR5 NUMA node closest to that GPU. Through the two \texttt{chunk\_size} variables, the ratio between HBM3 data and LPDDR5 data can be steered to maximize HBM usage, something that is not possible with plain \texttt{malloc()} and the \texttt{numactl --interleave} approach.

To evaluate the different possible approaches, we run a $N=33$ qubit benchmark with Hadamard gates and 11 passes requiring \qty{128}{\giga\byte} of memory on one GH200. Experiments with \texttt{malloc()} provided very poor performance\footnote{Conducted experiments have been canceled after exceeding
a runtime of several minutes.}. Monitoring GPU memory usage indicated that this was because Unified Memory Driver migration heuristics
caused frequent page migrations between LPDDR5 and HBM3 in the given memory oversubscription scenario. The same happened when combining
\texttt{malloc()} with \texttt{numactl --interleave}, and also when using the Unified Memory Data Usage Hints to explicitly stripe. This unexpected effect is
currently being investigated by the Unified Memory team at NVIDIA. We could work around the behavior by disabling access-counter-based memory
migration, as described in the CUDA 12.4 release notes~\cite{CUDA124RNGeneralCUDASec}.
While using Data Usage Hints, we experimentally identified a $4/11$ ratio of LPDDR5 to HBM3 data as the best-performing configuration. This closely approximates the theoretical optimum of $(128 - 96)/96 = 1/3$, while reserving sufficient space for auxiliary data structures.

\begin{table}[ht]
\begin{center}
\caption{Comparison of memory over-subscription strategies on one GH200 superchip.}
\label{tab:mem_oversubscription}
\resizebox{\columnwidth}{!}{
\begin{tabular}{lcc} 
\toprule
Strategy & Runtime (s) & Speedup (\emph{rel}) \\
\midrule
\circled[1pt]{1} \texttt{malloc()} & 446.67 & 1 \\
\circled[1pt]{2} \texttt{numactl} & 160.00 & 2.8 \\
\circled[1pt]{3} \texttt{numa\_alloc\_onnode()} & 101.57 & 4.4 \\
\circled[1pt]{4} Data Usage Hints & 71.70 & 6.2 \\
\circled[1pt]{5} Explicit Data Copies & 53.87 & 8.3 \\
\bottomrule
\end{tabular}}
\end{center}
\end{table}

Already with additional NUMA-related knowledge as provided with \texttt{numactl}, significant performance improvements can be gained.
The use of Data Usage Hints offers further gains with minimal programming effort and is particularly effective for applications with complex data patterns or in prototyping and exploratory scenarios.

The last, fifth option is to use explicit data movement functions to transfer data between HBM3 and LPDDR5 memory.
A library of such  explicit memory transfer functions is provided by the \emph{Host State Vector Migration} feature of NVIDIA cuQuantum 
but details of the implementation or performance have not been published.

JUQCS-50 surpasses the basic use of explicit data movement functions by integrating asynchronous CUDA copy functions with a look-ahead analyzer of the input gate sequence, enabling the automated, dynamic optimization of data movements on-the-fly.

The performance of the five approaches is compared in \autoref{tab:mem_oversubscription} (access-counter-based memory migrations disabled).
To enhance performance and ensure adaptability to applications requiring storage for multiple state vectors, we have opted to utilize explicit memory transfers and on-the-fly optimization of data movements.

\subsection{Inter-GH200 communication}
We use standard CUDA-aware MPI library calls to exchange data between different GH200s and nodes and adopt the technique of Ref.~\cite{RAED07x}, which minimizes the amount of data transfers automatically. {\color{black}A detailed description of this technique (which is different from the ShareMem approach of HyQuas~\cite{Zhang2021}), as well as  state vector distribution across GPUs and CPUs and the distinction between local and non-local qubits, can be found in our earlier works~\cite{RAED07x,RAED19a,Willsch2022} (Fig.~1 of~\cite{Willsch2022} applies to both JUWELS-Booster and JUPITER).}

Exploiting the asynchronous stream feature provided by CUDA, packing and unpacking data during the send and receive operations of MPI reduces the time to perform the inter-GH200 communication by approximately \qty{10}{\percent}.

In principle, the number of inter-GH200 data exchanges could be further reduced through independent preprocessing of the gate sequence, an approach external to JUQCS (see also \autoref{sec:adder}). As this represents a nontrivial challenge, it lies beyond the scope of the present work and is planned for future investigation. {\color{black}Other approaches for communication optimization, such as discussed in  Refs.~\cite{Jiao2023,Xu2024}, might be beneficial.}

\begin{table}[ht]
    \centering\renewcommand{\arraystretch}{1.3}
    \caption{\label{tab:jupiter}Specifications JUPITER Booster}
    \resizebox{\columnwidth}{!}{
    \begin{tabular}{lc}
    \toprule
        \textbf{Quantity} &\textbf{JUPITER Booster}\\
        \midrule
        Number of nodes & \num{5884} \\
        Number of GH200 & \num{23536} \\
        TDP per GH200 &\qty{680}{\watt} \\
        Total CPU LPDDR5 memory  & \qty{2880}{\tera\byte} \\
        Total GPU HBM3 memory  & \qty{2304}{\tera\byte}\\
        Interconnect Type & InfiniBand NDR200   \\
        Node Injection Bandwidth & $4\times\qty{200}{\giga\bit\per\second}$ \\
        Network Topology & Dragonfly+ \\
        Top500 Listing & Jun. 2025\\
        Rmax  & $\sim$\qty{1001}{\peta\floppersec} \\
        Rpeak  &$\sim$\qty{1300}{\peta\floppersec} \\
        \bottomrule
    \end{tabular}}
\end{table}

\subsection{Byte-encoding of the state vector elements}
\label{sec:byteencoding}
In earlier work~\cite{RAED19a}, we explored various ways to
encode the complex values of the state vector with less than two double-precision numbers ($2\times\qty{64}{\bit}$). An adaptive
encoding scheme that has been found to perform well 
is based on the polar representation
of a complex number  ($z=re^{i\theta}$) and uses an update strategy to adaptively and automatically tune the encoding/decoding scheme to the particular quantum circuit being executed.
A key feature of the scheme is its capacity to retain algorithmic precision to FP64-levels for some simulations, like the {\color{black}Hadamard benchmarks} presented in this work.
Obviously, using two ($2\times\qty{8}{\bit}$) instead of sixteen bytes ($2\times\qty{64}{\bit}$) to store each of the complex-valued elements of the state vector reduces the amount of memory by a factor of eight.

Of course, this memory reduction comes at the expense of additional compute time incurred by on-the-fly encoding and decoding during the execution of the gate. For instance, running the gate sequence for the $N=32$ qubit case (which fits in the HBM3 memory of a single GH200) takes \qty{5.74}{\second} and \qty{4.58}{\second} of computation time in byte-encoded and FP64 modes, respectively. {\color{black}For $N = 33$ qubits, however, using the memory of one GH200 superchip (HBM and LPDDR memory), in which only inter-superchip data is transferred, the byte-encoded code finishes the Hadamard benchmark in 13.5~s, whereas the FP64 finishes in 23.0~s, reflecting the fact that the amount of data transferred in the former is eight times smaller than in the latter.}

In general, the amount of additional compute complexity depends on the gate and varies from very little in the case of the X or CNOT gate to a factor of two to three (depending on the hardware) for gates such as the Hadamard gate~\cite{RAED19a}.
For the purpose of this work, the gate operations have been implemented as GPU device kernels. {\color{black}We emphasize that the adaptive encoding/decoding is performed on-the-fly at runtime.} 

\subsection{Encoding Choices and Their Tradeoffs}

{\color{black}
Predicting exactly which gate sequences will suffer significant accuracy loss under byte-encoding remains challenging. Generally, applications sensitive to high phase resolution or those spanning large amplitude ranges are most vulnerable. Ultimately, empirical validation is necessary for specific circuits; for example, Ref.~\cite{RAED19a} observed no significant deterioration in a 48-qubit implementation of Shor’s algorithm, and we report similar stability for a 50-qubit application in \autoref{sec:adder}.

While 50-qubit simulations on JUPITER currently necessitate byte-encoding to fit within memory constraints, JUQCS-50 is designed for versatility. It can utilize both HBM3 and LPDDR5 memory in FP32 or FP64 precision (see \autoref{tab:last} for a comparison of these modes). Specifically, using FP32 precision with LPDDR5 memory allows for the simulation of large circuits up to 48 qubits without noticeable loss of fidelity~\cite{MontanezBarrera2026}.

\begin{table}[ht]
    \centering\renewcommand{\arraystretch}{1.3}
    \caption{\label{tab:last}%
    {\color{black}The maximum number of qubits and a comparison of elapsed times (ET) in seconds and the number of GPUs (in parentheses) for running benchmarks on JUPITER Booster in the three different modes (\textbf{B}yte \textbf{E}ncoding, \qty{32}{\bit}, \qty{64}{\bit}).
    The second row shows how the elapsed times depend on the number of GPUs used, the third how the times change if number of GPUs is kept constant; in both cases, the number of qubits is kept constant.
    }}
\footnotesize
\begin{tabular}{lllll}
\toprule
&$N$&BE&FP32&FP64\\
\midrule
Maximum $N$& & 50& 48&47\\
ET / \unit{\second} (\#GPUs)&45& 223 (512) &  115 (2048) & 124 (4096) \\
ET / \unit{\second} (\#GPUs)&45& 43 (4096) &  67 (4096) & 124 (4096) \\
\bottomrule
\end{tabular}
\end{table}

As summarized in \autoref{tab:last}, the available execution modes reveal a clear trade‑off: FP32 typically offers the best balance of speed and accuracy, whereas byte‑encoding becomes essential for larger problems when reduced precision allows.}

\section{Performance criteria}
\label{sec:perfmeas}

The quantum computer simulator used in this work, JUQCS-50, was already introduced above. The base program, JUQCS, looks back on a long history of performance-conscious evaluations~\cite{RAED07x,RAED19a} and is used for a variety of scientific applications~\cite{RAED19a,Willsch2019}.
To evaluate the computational efficiency of JUQCS-50, we consider several performance metrics that capture both user-level experience and system-level behavior.

The most relevant measure from the user's perspective is the total elapsed time to simulate a quantum circuit. This walltime includes not only the gate operations themselves but also the initialization of data structures, MPI setup, I/O, and other runtime overheads.

Complementary metrics provide deeper insights into system performance, see \ref{appA}. These include the pure computation time, MPI communication time, and the time spent transferring data between HBM3 and LPDDR5 memory within each GH200 superchip---along with the associated data volumes.

The NVIDIA Grace Hopper GH200 superchip is core to our work, combining a 72-core, ARM-based CPU (Grace) with a Hopper GPU in one package, connected through a unique \qty{900}{\giga\byte\per\second} bus (NVLink-C2C, \emph{Chip-to-Chip}). The platform offers cache coherency for all involved memories, even between different superchips. Together with the fast NVLink-C2C, this cache coherency enables our memory-oversubscribing method, allowing access to the \qty{120}{\giga\byte} LPDDR5 host memory and \qty{96}{\giga\byte} HBM3 device memory from both processors with good performance. The involved bandwidths can be seen in the node diagram \autoref{fig:jupiter-node-diagram} with four GH200s.

JUPITER is built from NVIDIA GH200 superchips and provides four of these superchips per node. 
Since the available power is allocated per superchip, it must be shared between the Grace CPU and the Hopper GPU. Investigating how different power distribution strategies impact the overall performance is an important direction for future work. JUPITER employs an InfiniBand NDR fabric with a Dragonfly+ topology~\cite{dragonfly} with 25 groups and up to 240 nodes in a group, offering $4\times\qty{200}{\giga\bit\per\second}$ injection bandwidth per node. An overview of some defining system parameters is given in \autoref{tab:jupiter}.

JUQCS-50 is implemented in modern Fortran and utilizes CUDA Fortran for GPU acceleration through the NVIDIA HPC SDK toolkit, version 25.5, and GPU-aware OpenMPI or ParaStationMPI on JUPITER. All results reported in this work have been obtained with the OpenMPI version.

As previously discussed, the memory required to simulate an $N$-qubit universal quantum computer grows exponentially with $N$. This exponential growth imposes a hard limit on the number of qubits that can be simulated, namely $N=50$ on JUPITER. While the theoretical exponential scaling of computation time with $N$ can be almost entirely mitigated through massive parallelization~\cite{RAED07x,RAED19a}, as discussed in the next section, memory---rather than elapsed time---becomes the primary limiting factor for most applications. As a result, weak-scaling performance, as a function of $N$, serves as the key indicator of performance. For completeness, we also present strong-scaling data for $N=40$ qubits, where the number of GPUs increases from $4$ to $128$, with four GPUs per node.

\section{Performance Results}\label{sec:PerfResults}

In this section, we delve into the performance results obtained for JUQCS-50, focusing on the key metrics identified earlier. For clarity and completeness, detailed performance tables are provided in \ref{appA}.

\subsection{Quantum Circuit used for Benchmarking}

\begin{figure}[ht]
    \centering
    \includegraphics[scale=0.69]{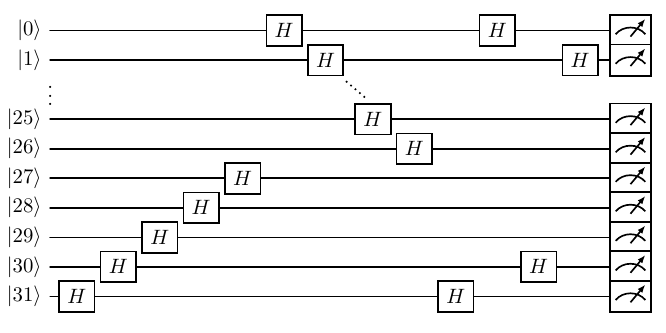}
    \caption{Graphical representation of the benchmark circuit \autoref{sequence}
    for the case of $N=32$ qubits. Operations proceed from left to right. Each application of a Hadamard gate ($H$) changes all the elements of the state vector. The rightmost symbol represents the simultaneous measurement of all three components of the Pauli-spin matrices representing a qubit, as performed by JUQCS-50. The initial state vector has all qubits in state zero.}
    \label{fig:benchmarkcircuit}
\end{figure}

To validate the results of byte-encoded simulations, it is expedient to use a gate sequence that does not suffer from precision loss due to this encoding.
One such sequence consists of Hadamard gates~\cite{Nielsen2010} followed by the simultaneous measurement of the three components {\color{black}(equivalent to measuring in the $X$, $Y$, $Z$ bases)} of each qubit (which is only possible in simulation{\color{black}, see~\cite{Gokhale2020}}). 
Specifically, we use
\begin{align}
\hbox{Circuit}=M\,&H_1\,H_{N-2}\,H_0\,H_{N-1}\,H_{N-6}\,H_{N-5}\,\ldots
\nonumber \\
&\ldots\,H_0\,H_{N-5}\,\ldots\,H_{N-1}
\;,
\label{sequence}
\end{align}
where $H_i$ denotes the Hadamard operation on qubit $i$ and $M$ represents the measurement of all qubits, see \autoref{fig:benchmarkcircuit} for a graphical representation.
This sequence keeps a reasonable balance between computational workload and MPI-based communication overhead
as $N$ is varied.
Note that the sequence \autoref{sequence} is to be read from right to left.

\subsection{Elapsed and Computation Times}
\autoref{tab1a} lists the elapsed and compute times for JUQCS-50 executing a sequence of Hadamard gates on JUPITER in adaptive byte-encoded mode, using HBM3 and LPDDR5 memory, and on-the-fly optimization of data exchange, performing computation on the GPU only.
In this mode, due to the intrinsic factor of two that is characteristic of quantum computers,
the largest quantum computer that can be simulated on JUPITER (\num{16384} GH200 superchips) contains $50$ qubits.
In the case of a sequence of Hadamard gates followed by a measurement of all qubits, the use of the adaptive byte-encoding does not cause a loss of precision; that is, the final data are FP64-accurate by design, providing a nontrivial validation of the code.

\begin{figure}[ht]
    \centering
    \includegraphics[width=\linewidth]{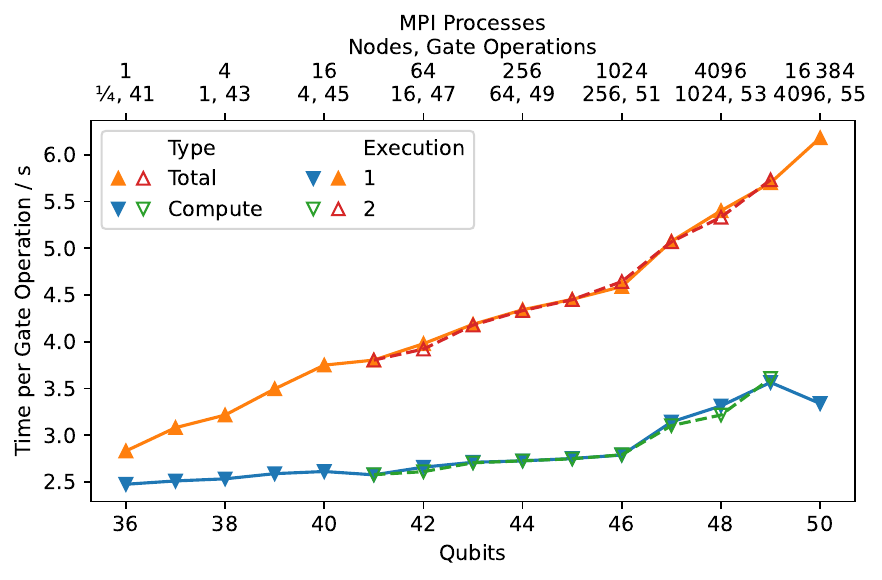}
    \caption{{\color{black}Total and compute elapsed time per gate operation for executing the circuit shown in \autoref{fig:benchmarkcircuit} on JUPITER as a function of the number of qubits (weak scaling). The first dataset (labeled “1”; see \autoref{tab1a}) was obtained with JUQCS running with exclusive access to the JUPITER supercomputer. The second dataset (labeled “2”) was collected on a different day, when other applications were running concurrently on JUPITER. Lines are shown as guides to the eye only.}}
    \label{fig:gate-op}
\end{figure}

\autoref{fig:gate-op} presents the total elapsed time per gate operation, as well as the computation time per gate, across a range of qubit counts for two distinct sets of runs. One of the most striking conclusions is the apparent high reproducibility of the data. {\color{black}The timings reported here are statistically significant. For the largest runs, the deviation from the expected trend is primarily due to external network activity caused by concurrent jobs on the system shared by many users. In addition, in byte-encoded mode, the reported compute time includes a small fraction of MPI-related overhead.}

Notably, the computation times remain nearly constant as the number of nodes (qubits) increases from \numrange{1}{256} (\numrange{36}{46}), while the total elapsed times reveal that the impact of network communication grows approximately linearly rather than exponentially with the number of qubits.

The computation time increases a little faster than theoretically expected. 
Based on the data in \autoref{tab1a}, and taking the $N=37$ case as a reference, we can estimate the expected computation time for the $N=46$ case under ideal parallelism as $\qty{106}{\second}\times51/42=\qty{129}{\second}$. This estimate is close to but still lower than the actual measured time of \qty{142}{\second}. The discrepancy arises from a combination of factors: interruptions in the byte decoding-encoding process due to global MPI communication (which is not accounted for in the reported MPI time), variability in GPU computation speeds, and network congestion.
These factors become more important for $N>46$, resulting in a notable change in the slopes of the lines through both the computation and elapsed times. At 47 qubits, executed on 512 nodes, the borders of one single Dragonfly group with 240 nodes are safely surpassed, and a majority of communication happens through the tapered network between the groups, rather than inside them.

\begin{figure}[ht]
    \centering
        \includegraphics[width=\linewidth]{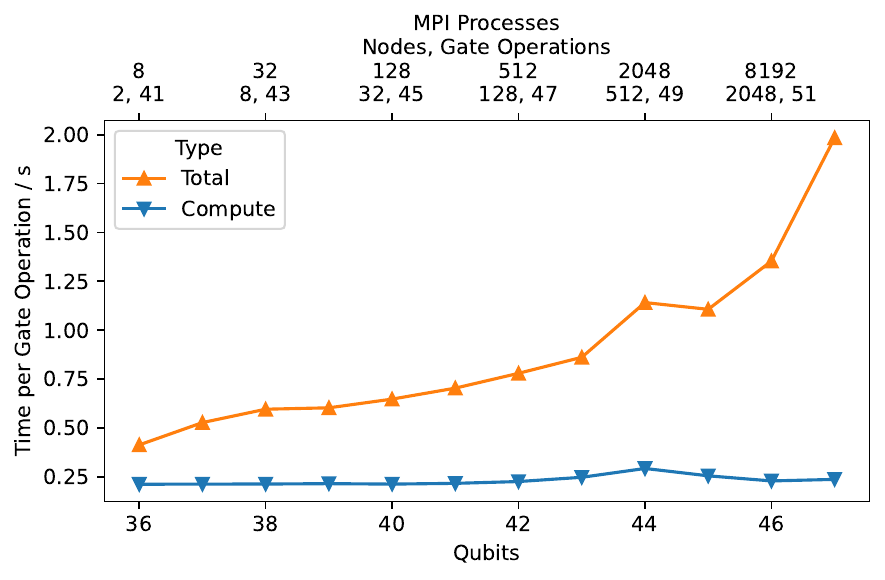}
    \caption{Total elapsed and compute times per gate operation for the range of qubits simulated on JUPITER in FP32 mode without using the LPDDR5 memory as an extension (weak scaling).
    When comparing to \autoref{fig:gate-op}, note the difference in scale of the $y$-axis 
    and keep in mind that the number of GPUs used is eight times larger, see \autoref{tabFP32}.  
    Lines are guides to the eye only.}
    \label{figFP32}
\end{figure}

To examine the combined impact of global MPI communication,
integral to the byte decoding–encoding process,
variability in GPU computation speeds, and network congestion, as well as to showcase the versatility of JUQCS-50, we conducted benchmarks in FP32 mode. 
In this configuration, state vector coefficients are stored in FP32 format, 
while all arithmetic operations are performed in FP64. 
Additionally, LPDDR5 was not used as a CPU memory extension. \autoref{figFP32} illustrates the computation time and total elapsed time per gate, 
with further details provided in \autoref{tabFP32}. 
The most striking observation is that, up to the maximum number of qubits supported by JUPITER in this mode,
namely $N=47$
and aside from a minor anomaly at $N=44$, the computation time remains nearly constant. 
This indicates near-perfect weak scaling behavior across the qubit range from 36 to 47.

As expected, the total elapsed time increases with $N$, driven by the growing volume of data exchanged 
and compounded by network congestion,
particularly since these benchmarks were executed concurrently with other users' workloads {\color{black}(all major MPI communication is of the blocking type, and the columns GPU-CPU time and data in the tables provide the necessary information about the use of asynchronous APIs (cudaMemcpyAsync))}. 

{\color{black}According to theory, the computational effort is expected to grow as $2^{N-36}$. Fitting the data (not shown) for elapsed times to $a+2^{b(N-36)}$ yields $a=-0.67$ and $b=0.11$. The value of $b$ is significantly smaller than the theoretically expected value of one, a manifestation of a high degree of parallelism of JUQCS. A slightly better fit (not shown) is obtained by using $a'+b'(N-36)^2$ with $a'=0.54$ and $b'=0.001$, suggesting that the data does not allow making conclusive assertions about the trend.}

We conclude that the change in slope observed for $N>46$ in \autoref{fig:gate-op} 
may be attributed to the increased computational burden of the byte decoding-encoding process, 
which demands substantially more arithmetic operations,  intermittently disrupted by global MPI communication, the topology,
and the fact that the FP32 benchmark did not use the LPDDR5 memory of the CPU as a memory extension.

Overall, the computation time demonstrates excellent weak scaling behavior. The impact of transmitting massive volumes of data across the network manifests as a relatively mild dependence on $N$ in the weak scaling of the total elapsed time, a point that will be examined in more detail later.

For the 36-qubit case, the elapsed times in \autoref{tab1a} and \autoref{tab3a} show that the FP64 version takes about \qty{54}{\second} to complete the task by using eight GH200 superchips, whereas the byte-encoded version accomplished the same task in about \qty{116}{\second} using only one GH200 superchip.
Put differently, the byte-encoded version is about a factor of $8\times54/(1\times116)\approx3.7$ times more efficient {\color{black} (see \autoref{tab1a})}.

Due to the communication making up a larger part of the elapsed time as the number of qubits increases, for the largest problem that can be run on JUPITER in FP64 mode, the gain in efficiency, due to using byte-encoding, is a factor $\num{16384}\times185/(2048\times264)\approx5.6$  {\color{black} (see \autoref{tab1a} and \autoref{tab3a})}.

{\color{black}To put the performance of the GH200-based system in perspective: setting the earlier world record~\cite{RAED19a} for $N=48$ qubits using byte-encoding for Hadamard benchmarks took the K computer \qty{3102}{\second} and the TaihuLight system (without using accelerators) \qty{8548}{\second}. For comparison, we compute the elapsed time/gate for the K computer and the JUPITER executing the $N=36$ Hadamard circuit under the exact same conditions, and find 47.03~s/gate for the former and 2.83~s/gate for the latter (see \autoref{tab1a}). This amounts to a performance increase by a factor of $16.6$ on the JUPITER system over the K computer.}

On the JUPITER supercomputer, the maximum number of qubits that can be simulated using JUQCS-50 varies by numerical precision and memory configuration. Specifically, in byte-encoded, FP32, and FP64 modes, the limits are 50 (49), 48 (47), and 47 (46) qubits, respectively, with {\color{black} (without)} LPDDR5 memory of the GH200. Using the largest FP64-simulatable problem (47 (46) qubits) as a reference, the corresponding benchmark execution times are 264 (101), 145 (103), and 185 (106) seconds for byte-encoded, FP32, and FP64 modes, respectively, utilizing 512 (512), 2048 (2048), and 4096 (4096) JUPITER nodes. These results demonstrate that for problems of large but not maximal size, JUQCS-50 enables optimization across numerical precision, runtime, and node count.

\subsection{Network Performance}
Regarding the data transfer rate between HBM3 and LPDDR5 memory in the same GH200 superchip, Tables~\ref{tab1a}--\ref{tab4a} show that this rate, which includes pre- and postprocessing of the buffers in GPU memory, is of the order of \qty{100}{\gibi\byte\per\second}.
The amount of data that is being exchanged within one GH200 superchip is (except for $N=36$) more than a factor of two larger than the data that is being sent and received by the same superchip over the interconnects (column GPU-GPU MPI data).
Because of the sequence-dependent pattern by which data is being sent over the network, a straightforward comparison of the transfer times is rather difficult.

\begin{figure}
    \centering
    \includegraphics[width=\linewidth]{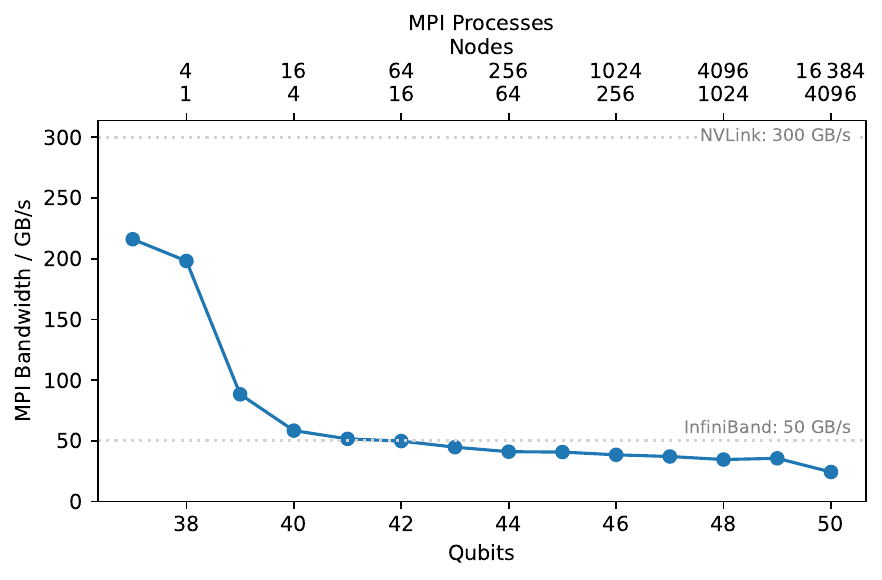}
    \caption{Measured communication bandwidth (in \unit{\gibi\byte\per\second}) as a function of qubits (bottom) and MPI processes/nodes (top) with the expected performance shown for JUPITER.  Horizontal lines indicate the limits of the interconnect bandwidths for comparison. The line through the data points is a guide to the eye only.}
    \label{fig:bw}
\end{figure}

For $N=50$ qubits, \qty{2048}{\tebi\byte} of data is required to represent the state vector. With each single-qubit gate that requires MPI communication, half of this data travels through the network. In total, for the sequences used, there are $36$ such network-active gate operations. The highly-optimized measurements of all the qubits require $13/14$ of all the data being exchanged.
Thus, in total, for
$N=50$, $ (36/2+13/14)\times 2048\approx \qty{38766}{\tebi\byte}$
traverses the network in about 100 seconds, a respectable amount of \unit{\tebi\byte} per second.

{\color{black} For $N>36$, $2^{(N-36+2)} = 2^{(N-34)}$  GPUs compete for network bandwidth to send and receive data. The  sharp drop in bandwidth for $38\le N \le 40$, signals the transition from intra-node-dominated, fast NVLink communication (see \autoref{fig:jupiter-node-diagram}) to the slower, inter-node-dominated network-mode of communication. Disregarding small 
fluctuations, the MPI times listed in \autoref{tab1a} fit well to the function $124.7/\sqrt{(N-36)}$ for $N \ge 40$. In this regime, the $N$ dependence shows up as a slowly decaying ($\approx{\cal O}(1/\sqrt{N})$) measured bandwidth (MPI data/MPI time; data taken from \autoref{tab1a}-\autoref{tab4a}) depicted in \autoref{fig:bw}. }

\subsection{Strong Scaling}

\begin{figure}[ht]
    \centering
    \includegraphics[width=\linewidth]{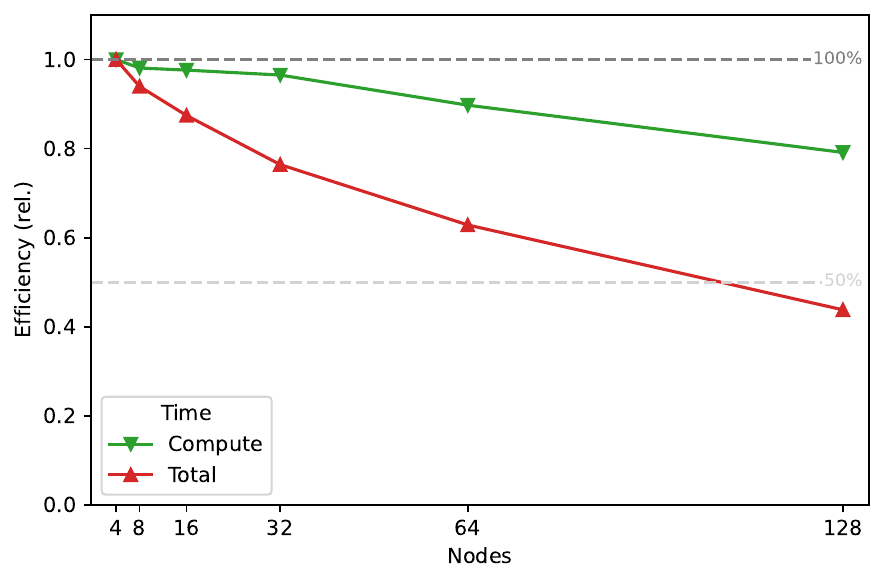}
    \caption{Relative efficiency measured (strong scaling) for 40 qubits over a range of MPI processes/nodes. Shown is the relative time with respect to the elapsed time on four nodes (16 MPI processes, $N=40$ qubits). The lines through the data points are guides to the eye only.}
    \label{fig:strong-rel}
\end{figure}

Strong scaling has no real-world relevance from the viewpoint of quantum processing hardware
because a quantum information processor having $K$ qubits has no way to distribute the execution over several units.
However, from the viewpoint of benchmarking and simulation, it is interesting to study the strong scaling behavior of the algorithm and hardware.
\autoref{tab4a} collects the data for fixed $N=40$ and
an increasing number of GPUs, always for the same quantum circuit, also see \autoref{fig:strong-rel}.

The strong scaling data shows that the computation times and GPU-CPU communication times scale close to perfect, but, as expected, the total elapsed times do not, due to the necessary communication among nodes.
As the number of MPI processes doubles, the GPUs have to exchange less data (by a factor of two) but they also have to compete with more GPUs (by a factor of two)
for network bandwidth.
From \autoref{tab4a}, it follows that the measured time per GiB is \numlist[list-units = single]{0.017;0.020;0.022;0.027;0.033;0.047} 
for $16,\ldots,512$ MPI processes, respectively.
As the number of MPI processes increases, there is an initial, approximately linear increase of the time per \unit{\gibi\byte}. Then, starting from 128 MPI processes, the time per \unit{\gibi\byte} is dominated by the slower inter-node network and the measured time per \unit{\gibi\byte} saturates at approximately \qty{0.047}{\second\per\gibi\byte} or, equivalently, a bandwidth of \qty{21}{\gibi\byte\per\second}, in rough agreement with the bandwidth (\qty{24}{\gibi\byte\per\second}, see \autoref{tab4a}) for a large number of qubits.
As there is the intrinsic limitation to powers of two, there is no way to determine accurately the number of MPI processes at which the saturation sets in.

\subsection{Hopper GPU  versus  ARM CPU}

Although not directly related to the use of the heterogeneous CPU-GPU architecture of the GH200 superchip,
benchmarking the CPU and GPU components separately is valuable from the standpoint of JUQCS-50 code portability.
\autoref{tab7} presents data enabling such a comparison across the three precision modes
currently supported by JUQCS-50.
From this data we conclude that (i) storing the state vector in FP32 precision yields the fastest computation
times for both CPU and GPU and (ii) MPI communication via the GPUs is faster than via the CPUs (which is to be expected
from the design of the GH200 chip), and (iii) depending on the precision, the Hopper GPU accomplished the task a factor of
$\approx$ 3 to 4 faster than the CPU. 
This is partially due to the fact that with JUQCS-50, the GPU computes faster than it can retrieve data from its HBM3 memory and, of course, this factor is also affected by the time used for MPI communication.

\begin{table}[h]
\centering
\caption{
Data obtained by running the benchmark circuit \autoref{sequence} on either the Hopper GPU or the ARM CPU (using 64 of the 72 cores) of the GH200 superchip for $N=40$ qubits and 256 MPI processes (64 nodes) {\color{black}with byte-encoding (BE), FP32, and FP64}.
}
\resizebox{\columnwidth}{!}{
\begin{tabular}{@{}ccccccc@{}}
\toprule
 elapsed & computation &    MPI    &  MPI  &  Compute  & precision of \\
      time (s) &   time (s)  & time (s) &  Data (GiB)  &  engine   & state vector   \\
\midrule
    16.14 &         7.24 &      6.08  &         111&   GPU &  BE    \\
    18.85 &         4.89 &     10.88  &         447&   GPU &  FP32  \\
    27.20 &         6.17 &     17.72  &         895&   GPU &  FP64  \\
\midrule
    47.61 &        25.76 &      9.67  &         111&   CPU &  BE    \\
    40.49 &        14.29 &     23.72  &         447&   CPU &  FP32  \\
    119.42 &        63.96 &     52.43  &         895&   CPU &  FP64  \\
\bottomrule
\end{tabular}}
\label{tab7}
\end{table}

\subsection{Summary}
Overall, \autoref{tab1a} and \autoref{fig:gate-op} show that the elapsed time increases approximately linearly (not exponentially) with the number of qubits $N$.
Recall that exponential scaling with the number of qubits $N$ is a prime characteristic of {\color{black}simulating} universal quantum computing. Therefore, the fact that a simulator such as JUQCS-50 can
simulate these computers in an elapsed time that grows approximately linearly with $N$ is significant. 

{\color{black}The simulation consumes \qty{58}{\watt\hour} energy for the 39 qubit case on one node, incorporating both CPU and GPU data. Beyond this single-node case, we observe a linear dependency between energy and number of nodes, with a robust slope of \qty{78}{\watt\hour}}.

Our focus in this work was on optimizing the data movement, both on-superchip and beyond, to enable simulation of 50 qubits for a first time. Certainly, further fine-tuning potential exists for the novel GH200 platforms.
In particular, the size of the buffers and the number of streams used for communication are unlikely to be optimal yet, and GPU device kernels that perform recursion can also be improved.

\section{Application: Adder Circuits}\label{sec:adder}

Sequences of Hadamard gates are well-suited for benchmarking purposes because:
(i) they are simple to implement and yield known outcomes,
(ii) their execution time is typically under 10 minutes,
(iii) they exert significant stress on the communication network, and
(iv) they maintain full precision without loss due to byte-encoding, which simplifies validation.

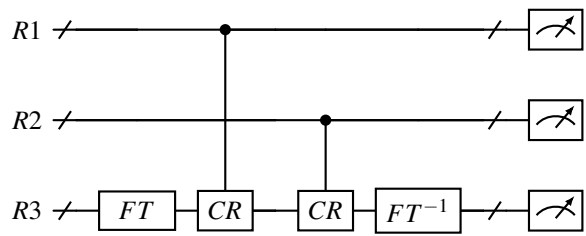
\begin{figure}[ht]
    \centering
    \begin{quantikz}[row sep={1.2cm,between origins}, column sep=0.3cm]
        \lstick{$R1$} & \qwbundle{} & \qw & \ctrl{2} & \qw & \qw & \qw & \qw & \qwbundle{} & \meter{} \\
        \lstick{$R2$} & \qwbundle{} & \qw & \qw & \qw & \ctrl{1} & \qw & \qw  & \qwbundle{}& \meter{} \\
        \lstick{$R3$} & \qwbundle{} & \gate[wires=2][1cm]{FT} & \gate{CR} & \qw & \gate{CR} & \gate{FT^{-1}} & \qw & \qwbundle{} & \meter{} \\
\end{quantikz}
    \caption{Structure of quantum circuit to add three $M$ bit integers encoded in quantum registers $R1$, $R2$, $R3$ consisting of $M$ qubits. The sum of the integers is returned in register $R3$, the qubits of other registers being untouched. By construction, integer addition is modulo $M$.
    FT: quantum circuit to perform a discrete Fourier transform~\cite{Nielsen2010}; CR: collection of controlled phase shifts. The right most symbol represents the simultaneous measurement of all three components of the Pauli-spin matrices representing a qubit, as performed by JUQCS-50.  The structure trivially generalizes to fewer and more registers encoding integers.}
    \label{fig:adder_schematic}
\end{figure}

\begin{figure*}[ht]
    \centering
    \includegraphics[scale=1]{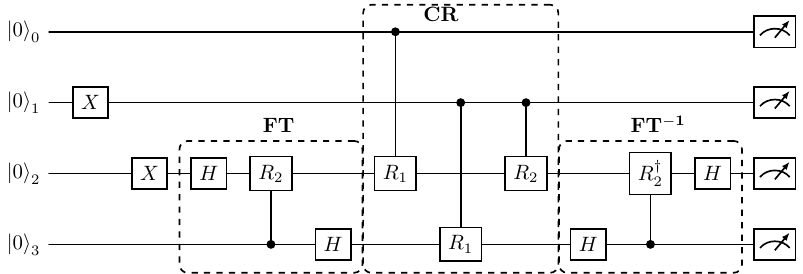}
    \caption{Realization of the quantum adder circuit shown in Fig.~\ref{fig:adder_schematic} for the case of two two-bit integers ($M=2$). The first two gates (denoted by $X$) are not part of the adder circuit but serve to encode the integers 1 and 2 in registers $R1$ and $R2$, respectively (see text).
    $X$: interchanges qubit state 0 and 1; $H$: Hadamard gate;
    $R_k$: controlled phase shift by $2\pi/2^k$. See Ref.~\cite{Nielsen2010} for a detailed description of these gates.
}
    \label{fig:2BitAdder}
\end{figure*}

Quantum algorithms that perform integer addition using quantum registers represent
another valuable class of quantum circuits.
(i) They are nontrivial in that they incorporate quantum Fourier transforms and controlled phase shifts,
both core components of foundational quantum algorithms such as
phase estimation~\cite{Nielsen2010} and Shor’s algorithm~\cite{Shor99,Willsch2023}.
(ii) The final output of the quantum computation, being the sum of the input integers, is easily validated.
(iii) These circuits can place a substantial load on the communication network,
offering a rigorous test of system performance.
(iv) When executed in byte-encoding mode, due to the presence of controlled phase shifts,
they may suffer a loss of numerical precision,
providing information about the effect of using byte-encoding
on the accuracy of the quantum computation.

A bird's-eye view of a quantum circuit to add (superpositions of) three integers stored
in the registers $R1$, $R2$, and $R3$, is shown in \autoref{fig:adder_schematic}.
Each of these registers is assumed to consist of $M$ qubits, enabling representing
integers in the range $[0,2^M-1]$.
Following Draper~\cite{DRAP00}, the idea is to perform a (quantum) Fourier transform (FT) on $R3$, then apply the controlled phase shifts ({\color{black}CR}) to accumulate the information stored in $R1$ and $R2$ in to the phases of the coefficients of the (partial) state vector in $R3$, and finally perform an inverse (quantum) Fourier transform (FT$^{-1}$) to transfer the information contained
in the phases back to integer representation.
The diagram in \autoref{fig:adder_schematic} illustrates the structure of a circuit that adds (superpositions of) integers stored in three registers.
However, the design generalizes straightforwardly to any number of registers.

\begin{table}[ht]
\caption{JUQCS-50 results for a 50-qubit quantum circuit designed to add two 25-bit integers (see text).
Only the 25 expectation values of the qubits in register $R2$ are shown (qubits are numbered starting from zero).
}
\begin{center}
\footnotesize
\begin{tabular}{cccc}
\toprule
Qubit&$\langle Q_x(i) \rangle$&$\langle Q_y(i)\rangle$ &$\langle Q_z(i)\rangle$ \\
\midrule
   24 & 0.50 & 0.50 & 1.00 \\
   25 & 0.50 & 0.52 & 1.00 \\
   26 & 0.51 & 0.50 & 1.00 \\
   27 & 0.50 & 0.50 & 1.00 \\
   28 & 0.50 & 0.50 & 1.00 \\
   29 & 0.50 & 0.50 & 1.00 \\
   30 & 0.50 & 0.50 & 1.00 \\
   31 & 0.50 & 0.50 & 1.00 \\
   32 & 0.50 & 0.50 & 1.00 \\
   33 & 0.50 & 0.50 & 1.00 \\
   34 & 0.50 & 0.50 & 1.00 \\
   35 & 0.50 & 0.50 & 1.00 \\
   36 & 0.50 & 0.50 & 1.00 \\
   37 & 0.50 & 0.50 & 1.00 \\
   38 & 0.50 & 0.50 & 1.00 \\
   39 & 0.50 & 0.50 & 1.00 \\
   40 & 0.50 & 0.50 & 1.00 \\
   41 & 0.50 & 0.50 & 1.00 \\
   42 & 0.50 & 0.50 & 1.00 \\
   43 & 0.50 & 0.50 & 1.00 \\
   44 & 0.50 & 0.50 & 1.00 \\
   45 & 0.50 & 0.50 & 1.00 \\
   46 & 0.50 & 0.50 & 1.00 \\
   47 & 0.48 & 0.49 & 1.00 \\
   48 & 0.50 & 0.50 & 1.00 \\
   49 & 0.50 & 0.50 & 1.00 \\
\bottomrule
\end{tabular}
\label{tabAdder}
\end{center}
\end{table}
As an illustration, \autoref{fig:2BitAdder} presents the quantum circuit for adding two integers
each one represented by a two-qubit register ($M=2$).
The $X$ gates flip the states of qubits 1 and 2 from $|0\rangle_1$ and $|0\rangle_2$ to $|1\rangle_1$ and $|1\rangle_2$, respectively.
Due of the structure of the FT used~\cite{Nielsen2010}, the most significant bit of the integer is stored
in the least significant bit of the register.
Therefore, after performing the $X$ gates, the state in R1 encodes the integer 1 and
the state in R2 encodes the integer 2.
Upon measurement, the state yields $\{0,1,1,1\}$, the last two bits encoding $1+2=3$.

As an illustration and also to scrutinize the effect of using byte-encoding on the final
result of a quantum computation, we use JUQCS-50 to execute a quantum circuit
for adding two 25-bit integers ($M=25$) with $R1$ encoding $\num{21346502}$ and $R2$ encoding $\num{12207929}$.
The quantum circuit contains $1001$ gates, which is too large to be displayed graphically on one page.
The integers have been chosen so that their binary sum (being $2^{25}-1$) results in a sequence of all ones.
Translated into qubit language, this implies that all the expectation values of the $z$-components of the qubits
($\langle Q_z(i)\rangle$)
are equal to one, allowing the correctness of the output to be visually confirmed at a glance.

\autoref{tabAdder} shows the results of the simulation.
It is immediately clear that the quantum circuit yields the correct values of $\langle Q_z(i)\rangle$.
Also clear is that due to the use of byte encoding,
some expectation values of the $x$- and $y$-components deviate from the exact result $1/2$.

It is worth noting that, unless proven otherwise, current quantum devices are unlikely to perform integer addition with the level of precision achieved by JUQCS-50 in its byte-encoded mode~\cite{MICH17b}.

The quantum adder circuit also provides a nice example to explore how much can be gained by optimizing
the circuit with respect to the amount of MPI communication.
In its generic form shown in \autoref{fig:adder_schematic}, it took JUPITER \qty{1996}{\second} (\qty{180}{\second} MPI GPU-GPU time, \qty{720}{\second} GPU-CPU time) to execute the quantum circuit.
Simply interchanging the roles or $R1$ and $R2$ by relabeling the qubits (an operation that is part of JUQCS-50's
instruction repertoire), reduces this time to \qty{1193}{\second} (\qty{69}{\second} MPI GPU-GPU time, \qty{335}{\second} GPU-CPU time).
This substantial performance improvement is solely attributed to the reduction in
communication.

More broadly, it is evident that strategically relabeling qubits to minimize inter-GPU communication
can significantly decrease the total execution time of simulating a quantum circuit.
Optimizing the quantum circuit in this regard can be performed independently of JUQCS-50 and
is therefore beyond the scope of the present study, but important future research.

\section{Conclusion}\label{sec:Summary}

JUQCS-50 efficiently utilizes the available hardware, CPUs, GPUs, and a combination of both, by exploiting the architectural strengths of modern systems like the GH200 superchips. 
{\color{black}Relative to the works reported in~\cite{RAED07x,RAED19a,Willsch2022}, 
the main technical innovations introduced in JUQCS-50 are:
\begin{enumerate}
\item
Explicit management of CPU--GPU memory transfers, 
including on-the-fly optimization of data movement between GPU and CPU memory within each node, 
enabling efficient use of the heterogeneous memory hierarchy.
\item
Use of CUDA-accelerated packing and unpacking of amplitude data during MPI send/receive operations,
significantly reducing communication overhead between nodes.
\item
Support for storing state-vector amplitudes in FP32 format, lowering memory requirements and communication volume 
while maintaining sufficient numerical accuracy for universal-circuit simulation.
\item
GPU kernels for byte-encoded operations.
\end{enumerate} 
}

These innovations not only enable the aforementioned large-scale realistic simulations of user-defined circuits but, at the same time, combined with its ease of deployment and operation, JUQCS-50 can continuously and controllably stress both computational units and the network over defined periods, making it an ideal and realistic benchmark user application.

From the data presented in Tables~\ref{tab1a}--\ref{tab3a}, it follows that, to a good approximation, the elapsed time increases approximately linearly with the number of qubits $N$, rather than exponentially. In other words, when combined with massively parallel computers, the universal quantum computer simulator JUQCS-50 overcomes the exponential scaling characteristic of gate-based quantum computers.

The advanced JUQCS-50, the newly developed version of JUQCS, will empower researchers to
1) perform exact simulations of arbitrary quantum circuits with varying degrees of accuracy, still far better than offered by current, state-of-the-art quantum computing hardware;
2) study the effects of noise and errors on quantum algorithms;
3) perform simulations of quantum annealing and quantum spin-dynamics
for a wide variety of model Hamiltonians over a time span and with an accuracy
that is not within reach of state-of-the-art quantum computing hardware;
4) expand simulations to larger quantum systems while keeping simulation times low;
5) benchmark (super)computers~\cite{RAED07x,RAED19a,Willsch2019,Herten2024}.

JUQCS-50 is currently being integrated into JUNIQ~\cite{JUNIQ}, 
the J\"ulich UNified Infrastructure for Quantum Computing, providing science and industry access to state-of-the-art quantum computing emulators and devices.
With JUPITER operational, JUQCS-50 will enable the study of problems up to 500 to 1000 times larger than those currently handled by other simulators (as discussed in \autoref{STATE}), thereby unlocking a new range of universal quantum computing applications yet to be explored.
More specifically, JUQCS-50 will allow researchers to run applications such as variational quantum eigensolvers (VQE) ~\cite{Peruzzo2014,McClean2016,Cerezo2021},
quantum approximate optimization algorithms (QAOA)~\cite{Farhi2014,RAED19a,Willsch2020,Zhou2020,Lotshaw2022,MontanezBarrera2026},
and quantum annealing~\cite{Vyas2026} 
with an accuracy that is beyond the reach of state-of-the-art quantum computer hardware for systems with up to 50 qubits.

\section*{Acknowledgments}
\addcontentsline{toc}{section}{Acknowledgments}
\footnotesize

We acknowledge support from the following entities: 
The Ministry of Culture and Science of the State of North Rhine-Westphalia (MKW-NRW) for the project EPIQ;
MKW-NRW together with EuroHPC JU, the German Federal Ministry of Education and Research (BMBF) for the project JUNIQ.
This project received access to the JUPITER supercomputer, which is funded by the EuroHPC Joint Undertaking, the German Federal Ministry of Research, Technology and Space, and the Ministry of Culture and Science of the German state of North Rhine-Westphalia, through the JUPITER Research and Early Access Program (JUREAP).
We thank the Swiss National Supercomputing Center CSCS for providing access to the Alps supercomputer, which was essential to prepare our workload for JUPITER.

\medskip
{\bf CRediT authorship contribution statement}

Hans De Raedt: Conceptualization, software, investigation, writing – original draft, writing– review \& editing; Jiri Kraus: Conceptualization, software, writing – original draft, writing– review \& editing; Andreas Herten: investigation, resources, visualization, writing – original draft, writing – review \& editing; Vrinda Mehta: investigation, validation, writing – original draft, writing – review \& editing; Mathis Bode: resources, writing – review \& editing; Markus Hrywniak: Software, writing – review \& editing; Kristel Michielsen: Writing – review \& editing, funding acquisition; 
Thomas Lippert: Writing – review \& editing, funding acquisition.

\medskip
{\bf Data availability}
Data will be made available on reasonable request.

\medskip
{\bf Declaration of competing interest}
The authors declare that they have no known competing financial interests or personal relationships that could have appeared to influence the work reported in this paper.

\medskip
{\bf Declaration of generative AI and AI-assisted technologies in the manuscript preparation process}

During the preparation of this work, the authors used ChatGPT, Gemini, and Copilot to enhance the language of the text. After using this tool, the authors reviewed and edited the content as needed and take full responsibility for the content of the published article.

\footnotesize
\bibliographystyle{elsarticle-num-names}
\bibliography{all26,references}

\clearpage
\appendix
\section{Raw performance data}\label{appA}
\begin{table}[H]
\captionsetup{width=\textwidth, justification=justified}
\makebox[\textwidth][c]{
   \parbox{\textwidth}{
        \caption{Elapsed and compute times for executing Hadamard gates~\cite{Nielsen2010} in a sequence designed to challenge
        both computation and communication on JUPITER, with JUQCS-50 employing adaptive byte-encoding,
        using both HBM3 and LPDDR5 memory and on-the-fly optimization of data exchange (weak scaling).
        The column GPU-GPU MPI lists the data sent and received by each of the GPUs (using CUDA-aware MPI). MPI time includes the time encoding and decoding buffers for transmitting these data.
        The columns 'GPU-CPU data' and 'count' specify the volume of data and the number of data exchanges between HBM3 and LPDDR5 memory. GPU-CPU time includes the time for writing and reading buffers in HBM3 memory.
        The column \#gate operations is the number of Hadamard operations plus the measurement of each of the qubits. 
        The simulation of a 50-qubit, universal quantum computer on JUPITER sets a new world record.
\label{tab1a}
}
        }
        }

\resizebox{\textwidth}{!}{
\begin{tabular}{@{}ccccccccccc@{}}
\toprule
Qubits& memory &     MPI   &  elapsed & computation &    \#gate    &   MPI    & GPU-GPU MPI  &  GPU-CPU  &  GPU-CPU   & GPU-CPU \\
      &  (GiB) & processes & time (s) &   time (s)  &  operations & time (s) &  Data (GiB)  &  time (s) & Data (GiB) &  count  \\
\midrule
    36 &     1026 &          1 &    116.13 &       101.54 &           41 &      0.00 &            0 &       11.60 &        1216 &       19 \\
    37 &     2052 &          2 &    129.43 &       105.51 &           42 &      2.37 &          512 &       17.99 &        2048 &       32 \\
    38 &     4104 &          4 &    138.31 &       108.99 &           43 &      4.20 &          832 &       20.23 &        2368 &       37 \\
    39 &     8208 &          8 &    153.83 &       113.96 &           44 &     12.68 &         1120 &       22.46 &        2688 &       42 \\
    40 &    16416 &         16 &    168.77 &       117.59 &           45 &     21.63 &         1264 &       24.54 &        2944 &       46 \\
    41 &    32832 &         32 &    175.00 &       118.55 &           46 &     24.64 &         1272 &       26.43 &        3136 &       49 \\
    42 &    65664 &         64 &    187.04 &       125.00 &           47 &     28.19 &         1404 &       28.51 &        3392 &       53 \\
    43 &   131328 &        128 &    200.97 &       130.16 &           48 &     34.25 &         1534 &       30.59 &        3648 &       57 \\
    44 &   262656 &        256 &    212.78 &       133.61 &           49 &     40.50 &         1663 &       32.64 &        3904 &       61 \\
    45 &   525312 &        512 &    222.71 &       137.59 &           50 &     43.94 &         1791 &       34.77 &        4160 &       65 \\
    46 &  1050624 &       1024 &    234.13 &       142.25 &           51 &     49.82 &         1919 &       36.81 &        4416 &       69 \\
    47 &  2101248 &       2048 &    263.86 &       163.43 &           52 &     55.15 &         2047 &       38.88 &        4672 &       73 \\
    48 &  4202496 &       4096 &    286.27 &       175.67 &           53 &     62.81 &         2175 &       40.94 &        4928 &       77 \\
    49 &  8404992 &       8192 &    307.83 &       192.56 &           54 &     64.65 &         2303 &       43.00 &        5184 &       81 \\
    50 & 16809984 &      16384 &    339.96 &       183.69 &           55 &     99.81 &         2431 &       45.08 &        5440 &       85 \\
\bottomrule
\end{tabular}}
\end{table}

\begin{table}[H]
\captionsetup{width=\textwidth, justification=justified}
\makebox[\textwidth][c]{%
  \parbox{\textwidth}{%
        \caption{Same as Table~\ref{tab1a} except that all calculations were
        performed with JUQCS-50 running in FP32 mode and without using the LPDDR5 memory as an extension (weak scaling) .\hfill
\label{tabFP32}
}
        }
    }
\resizebox{\textwidth}{!}{
\begin{tabular}{@{}ccccccccccc@{}}
\toprule
Qubits& memory &     MPI   &  elapsed & computation &    \#gate    &   MPI    & GPU-GPU MPI  &  GPU-CPU  &  GPU-CPU   & GPU-CPU \\
      &  (GiB) & processes & time (s) &   time (s)  &  operations & time (s) &  Data (GiB)  &  time (s) & Data (GiB) &  count  \\
\midrule
    36 &     1040  &         8   &   16.92  &        8.66 &           41 &      6.26  &         560 &        0.00 &           0 &        0  \\
    37 &     2080  &        16   &   22.13  &        8.91 &           42 &     10.88  &         632 &        0.00 &           0 &        0  \\
    38 &     4160  &        32   &   25.60  &        9.13 &           43 &     13.74  &         700 &        0.00 &           0 &        0  \\
    39 &     8320  &        64   &   26.51  &        9.43 &           44 &     14.29  &         766 &        0.00 &           0 &        0  \\
    40 &    16640  &       128   &   29.12  &        9.54 &           45 &     16.33  &         831 &        0.00 &           0 &        0  \\
    41 &    33280  &       256   &   32.35  &        9.93 &           46 &     19.14  &         895 &        0.00 &           0 &        0  \\
    42 &    66560  &       512   &   36.62  &       10.58 &           47 &     22.38  &         959 &        0.00 &           0 &        0  \\
    43 &   133120  &      1024   &   41.33  &       11.85 &           48 &     26.90  &        1023 &        0.00 &           0 &        0  \\
    44 &   266240  &      2048   &   55.92  &       14.31 &           49 &     37.87  &        1087 &        0.00 &           0 &        0  \\
    45 &   532480  &      4096   &   55.33  &       12.72 &           50 &     38.03  &        1151 &        0.00 &           0 &        0  \\
    46 &  1064960  &      8192   &   69.02  &       11.64 &           51 &     50.69  &        1215 &        0.00 &           0 &        0  \\
    47 &  2129920  &     16384   &  103.19  &       12.28 &           52 &     78.34  &        1279 &        0.00 &           0 &        0  \\
\bottomrule
\end{tabular}}
\end{table}

\clearpage

\begin{table}[h]
\captionsetup{width=\textwidth, justification=justified}
\makebox[\textwidth][c]{%
  \parbox{\textwidth}{%
        \caption{Same as Table~\ref{tab1a} except that all calculations were
        performed with JUQCS-50 running in FP64 mode (weak scaling).\hfill
\label{tab3a}
}
        }
    }
\resizebox{\textwidth}{!}{
\begin{tabular}{@{}ccccccccccc@{}}
\toprule
Qubits& memory &     MPI   &  elapsed & computation &    \#gate    &   MPI    & GPU-GPU MPI  &  GPU-CPU  &  GPU-CPU   & GPU-CPU \\
      &  (GiB) & processes & time (s) &   time (s)  &  operations & time (s) &  Data (GiB)  &  time (s) & Data (GiB) &  count  \\
\midrule
    36 &     1040   &        8 &     53.75 &        11.06   &         41 &     12.11 &         1120  &      25.24  &       2560    &    40 \\
    37 &     2080   &       16 &     65.17 &        12.60   &         42 &     22.12 &         1264  &      24.45  &       2816    &    44 \\
    38 &     4160   &       32 &     63.98 &        11.47   &         43 &     24.00 &         1272  &      23.04  &       3008    &    47 \\
    39 &     8320   &       64 &     70.22 &        12.10   &         44 &     27.65 &         1404  &      24.97  &       3264    &    51 \\
    40 &    16640   &      128 &     82.74 &        15.69   &         45 &     32.89 &         1534  &      27.47  &       3520    &    55 \\
    41 &    33280   &      256 &     82.71 &        12.81   &         46 &     35.04 &         1663  &      28.78  &       3776    &    59 \\
    42 &    66560   &      512 &     87.52 &        13.16   &         47 &     37.42 &         1791  &      30.65  &       4032    &    63 \\
    43 &   133120   &     1024 &     98.06 &        13.86   &         48 &     46.24 &         1919  &      32.54  &       4288    &    67 \\
    44 &   266240   &     2048 &    109.98 &        13.56   &         49 &     55.18 &         2047  &      34.46  &       4544    &    71 \\
    45 &   532480   &     4096 &    124.06 &        16.24   &         50 &     59.02 &         2175  &      36.36  &       4800    &    75 \\
    46 &  1064960   &     8192 &    150.05 &        18.61   &         51 &     65.45 &         2303  &      38.25  &       5056    &    79 \\
    47 &  2129920   &    16384 &    185.19 &        32.99   &         52 &     91.09 &         2431  &      40.18  &       5312    &    83 \\
\bottomrule
\end{tabular}}
\end{table}

\begin{table}[h]
\captionsetup{width=\textwidth, justification=justified}
\makebox[\textwidth][c]{%
  \parbox{\textwidth}{%
        \caption{
        Same as Table~\ref{tab1a} except that all calculations were
        performed for the same number of qubits $N=40$ (strong scaling).\hfill
\label{tab4a}
        }
        }
    }
\resizebox{\textwidth}{!}{
\begin{tabular}{@{}ccccccccccc@{}}
\toprule
Qubits& memory &     MPI   &  elapsed & computation &    \#gate    &   MPI    & GPU-GPU MPI  &  GPU-CPU  &  GPU-CPU   & GPU-CPU \\
      &  (GiB) & processes & time (s) &   time (s)  &  operations & time (s) &  Data (GiB)  &  time (s) & Data (GiB) &  count  \\
\midrule
    40 &    16416  &        16 &    166.90   &     115.75 &           45   &   21.62    &      1264   &     24.55      &   2944  &      46 \\
    40 &    16448  &        32 &     88.80   &      58.99 &           45   &   12.79    &       636   &     13.23      &   1568  &      49 \\
    40 &    16512  &        64 &     47.71   &      29.64 &           45   &    7.73    &       351   &      7.14      &    848  &      53 \\
    40 &    16640  &       128 &     27.30   &      14.99 &           45   &    5.27    &       191   &      3.83      &    456  &      57 \\
    40 &    16896  &       256 &     16.59   &       8.06 &           45   &    3.46    &       103   &      2.05      &    244  &      61 \\
    40 &    17408  &       512 &     11.90   &       4.57 &           45   &    2.61    &        55   &      1.09      &    130  &      65 \\
\bottomrule
\end{tabular}}
\end{table}
\end{document}